\documentclass[10pt]{article}
\usepackage{amsmath}
\usepackage{amsfonts}

\usepackage{mathrsfs,amsthm,amsmath,amssymb}
\usepackage{graphicx}
\usepackage{color}
\usepackage{url}

\theoremstyle{definition}
\newtheorem{Theorem}{Theorem}
\newtheorem{Corollary}{Corollary}

\newtheorem{Lemma}{Lemma}

\newcommand{\fpm}{{\mathbb F}_{p^m}}
\newcommand{\ftwon}{{\mathbb F}_{2^n}}
\newcommand{\ftwom}{{\mathbb F}_{2^m}}

\newcommand{\Proof}{\noindent\textbf{Proof.}~}
\newcommand{\done}{\hfill $\Box$ }

\newcommand{\ls}[1]
    {\dimen0=\fontdimen6\the\font\lineskip=#1\dimen0
     \advance\lineskip.5\fontdimen5\the\font
     \advance\lineskip-\dimen0
     \lineskiplimit=0.9\lineskip
     \baselineskip=\lineskip
     \advance\baselineskip\dimen0
     \normallineskip\lineskip\normallineskiplimit\lineskiplimit
     \normalbaselineskip\baselineskip
     \ignorespaces}

\begin{document}

\bibliographystyle{abbrv}

%\title{A class of cubic monomial  negabent functions}

\title{New Permutation Trinomials From Niho Exponents over Finite Fields with Even Characteristic}
\author{ Nian Li and Tor Helleseth
\thanks{The authors are with the Department of Informatics, University of Bergen,
 N-5020 Bergen, Norway. Email: Nian.Li@ii.uib.no;
Tor.Helleseth@ii.uib.no.}
}
\date{}
\maketitle
\ls{1.5}

\thispagestyle{plain} \setcounter{page}{1}

\begin{abstract}
 In this paper, a class of permutation trinomials of Niho type over finite fields with even characteristic is further investigated. New permutation trinomials from Niho exponents are obtained from linear fractional polynomials over finite fields, and it is shown that the presented results are the generalizations of some earlier works.
\end{abstract}

\section{Introduction}

Let $p$ be a prime and $\fpm$ denote the finite field with $p^m$ elements. A polynomial $f(x)\in\fpm[x]$ is called a permutation polynomial of $\fpm$ if the associated polynomial function $f: c\mapsto f(c)$ from $\fpm$ to $\fpm$ is a permutation of $\fpm$ \cite{Lidl-N}. Permutation polynomials over finite fields, which were first studied by Hermite \cite{Hermite} and Dickson \cite{Dickson}, have wide applications in the areas of mathematics and engineering such as coding theory, cryptography and combinatorial designs.

The construction of permutation polynomials with either a simple form or certain desired property is an interesting research problem. For some recent constructions of permutation polynomials with a simple form, the reader is referred to \cite{Ding-QWYY,Hou14,Hou15,Li-Qu-Chen,Tu-Zeng-Hu-Li,Tu-Zeng-Hu,Zieve-subgroup} and for some constructions of permutation polynomials with certain desired property, the reader is referred to \cite{BTT,Dobbertin,Dobbertin-Nihocase,LHT,Zeng,ZTT,ZZC} and the references therein for example. Permutation binomials and trinomials attract researchers' attention in recent years due to their simple algebraic form, and some nice works on them had been achieved in \cite{Ding-QWYY,Hou15,Tu-Zeng-Hu-Li,Zieve-subgroup}. However, currently only a small number of specific classes of permutation binomials and trinomials are described in the literature, which can be found in the survey paper \cite{Hou} on the developments of the constructions of permutation polynomials and four recent papers \cite{Ding-QWYY,LH,Li-Qu-Chen,ZZC} on permutation binomials and trinomials over finite fields.

In this paper, following the line of the works did in \cite{Tu-Zeng-Hu-Li,LH}, we further study the permutation trinomials from Niho exponents over finite fields with even characteristic. Precisely, let $n=2m$ be a positive integer and $p=2$, we consider the trinomials over the finite field $\ftwon$ of the form
\begin{eqnarray}\label{def-PP}
  f(x)=x+x^{s(2^m-1)+1}+x^{t(2^m-1)+1},
\end{eqnarray}
where $s,t$ are integers satisfying $1\leq s, t\leq 2^m$. For simplicity, if the integers $s, t$ are written as fractions or negative integers, then they should be interpreted as modulo $2^m+1$. For instance, $(s,t)=(\frac{1}{2},\frac{3}{4})=(2^{m-1}+1,2^{m-2}+1)$. The objective of this paper is then to find new pairs $(s,t)$ such that the trinomial $f(x)$ defined by \eqref{def-PP} is a permutation polynomial. Based on the analysis of certain equations over the finite field $\ftwon$, some new classes of permutation trinomials over $\ftwon$ with the form \eqref{def-PP} are constructed. The presented new classes of permutation trinomials of the form \eqref{def-PP} are determined by making use of the property of the linear fractional polynomials over the finite field $\ftwon$, which is a completely different approach, and it will be seen that many  permutation trinomials of the form \eqref{def-PP} can be obtained from them since their parameters on $(s,t)$ are flexible.

%The first new class of permutation trinomials of the form \eqref{def-PP} is obtained by using the same techniques as in \cite{LH}, i.e., by solving some low-degree equation over the finite field $\ftwon$.

\section{Preliminaries} %

Throughout this paper, let $n=2m$ and $\ftwon$ denote the finite field with $2^n$ elements.   A positive integer $d$ is called a {\em Niho exponent} with respect to the finite field $\ftwon$ if $d \equiv 2^j\pmod{2^m-1}$ for some nonnegative integer $j$. When $j=0$, the integer $d$ is then called a normalized Niho exponent. The Niho exponents were originally introduced by Niho \cite{Niho-PhD} who investigated the cross-correlation between an $m$-sequence and its $d$-decimation.

For simplicity, denote the conjugate of $x\in\ftwon$ over $\ftwom$ by $\overline{x}$, i.e., $\overline{x}=x^{2^m}$. The unit circle of $\ftwon$ is defined as follows:
\[U=\{x\in\ftwon: x\overline{x}=x^{2^m+1}=1\}.\]

Notice that the trinomial $f(x)$ defined by \eqref{def-PP} can be written as the form $x^rg(x^s)$ for some integers $r,s$ and $g(x)\in \ftwon[x]$. For this kind of polynomials, its permutation property had been characterized by the following lemma which was proved by Park and Lee in 2001 and reproved by Zieve in 2009.

 \begin{Lemma}(\cite{Park-Lee,Zieve-09}) \label{lem-Zieve}
  Let $p$ be a prime, $n$ a positive integer and $g(x)\in \mathbb{F}_{p^n}[x]$. If $d, s, r>0$ such that $p^n-1=ds$, then $x^rg(x^s)$ is a permutation over $\mathbb{F}_{p^n}$ if and only if
 \begin{enumerate}
   \item [1)] $\gcd(r,s)=1$, and
   \item [2)] $h(x)=x^rg(x)^s$ permutes the $d$-th roots of unity in $\mathbb{F}_{p^n}$.
 \end{enumerate}
\end{Lemma}

In the sequel, we will use the property of the linear fractional polynomial to prove our main results. For the linear fractional polynomial $\phi(x)=\frac{ax+b}{cx+d}$ with $a,b,c,d\in\ftwon$ and $ad-bc\ne 0$, it can be readily verified that  $\phi(x)$ defines a permutation on $\ftwon\cup\{\infty\}$, where
\begin{eqnarray*}
 \frac{a}{0}=\infty \;{\rm if}\, a \ne 0; \;\;\frac{a\infty+b}{c\infty+d}=\frac{a}{c}\; {\rm if}\, c \ne 0; \;\; \frac{a\infty+b}{c\infty+d}=\infty \;{\rm if}\, c = 0\, {\rm and}\, ad \ne 0.
\end{eqnarray*}
Then, by utilizing the fact that the composition of any two linear fractional polynomial is still a linear fractional polynomial, we can prove

 \begin{Lemma} \label{lem-phi}
  Let $a\in U$ with $a^3\ne 1$ and $\phi(x)=\frac{ax+(a+1)}{(a+1)x+1}$. If $\phi(x)=x^{2^k}$ holds for some positive integer $k$. Then for any $i\ge 2$ we have
 \begin{eqnarray*}%\label{eq-2ik}
 x^{2^{ik}}=\left \{ \begin{array}{cl}
x, & \textrm{if } e_{i-1}^{2^k}+e_1+1=0, \\
\frac{e_ix+1}{x+e_i+1}, & \textrm{if } e_{i-1}^{2^k}+e_1+1\not=0,
\end{array}\right.
\end{eqnarray*}
\end{Lemma}
where $e_1=\frac{a}{a+1}$, $e_i=\infty$ if $e_{i-1}^{2^k}+e_1+1=0$ and otherwise $e_i=\phi(e_{i-1}^{2^k})$ for any $i\ge 2$.

\Proof Note that $\phi(x)=\frac{ax+(a+1)}{(a+1)x+1}=\frac{e_1x+1}{x+e_1+1}$, where $e_1=\frac{a}{a+1}$.
 Taking $2^k$-th power on both sides of $x^{2^k}=\phi(x)$ gives
 \begin{eqnarray*}
 x^{2^{2k}}=\frac{e_1^{2^k}x^{2^k}+1}{x^{2^k}+e_1^{2^k}+1}.
\end{eqnarray*}
Then, substituting $x^{2^k}$ by $\phi(x)=\frac{e_1x+1}{x+e_1+1}$, we can further obtain
 \begin{eqnarray*}
 x^{2^{2k}}=\frac{(e_1^{2^k+1}+1)x+e_1^{2^k}+e_1+1}{(e_1^{2^k}+e_1+1)x+e_1^{2^k+1}+e_1^{2^k}+e_1}=\frac{e_2x+1}{x+e_2+1}
\end{eqnarray*}
if $e_1^{2^k}+e_1+1\ne 0$, where $e_2=\frac{e_1^{2^k+1}+1}{(e_1^{2^k}+e_1+1)}=\phi(e_1^{2^k})$. If $e_1^{2^k}+e_1+1=0$, then $x^{2^{2k}}=\frac{(e_1^2+e_1+1)x}{e_1^2+e_1+1}=x$
since $e_1^2+e_1+1\not=0$. Otherwise we have $e_1^3=(\frac{a}{a+1})^3=1$, i.e., $a^2+a+1=0$ which implies that $a^3=1$, a contradiction with $a^3\ne 1$. By using this, we then can show by induction on $i$ that
 \begin{eqnarray}\label{eq-2ik}
 x^{2^{ik}}=\left \{ \begin{array}{cl}
x, & \textrm{if } e_{i-1}^{2^k}+e_1+1=0, \\
\frac{e_ix+1}{x+e_i+1}, & \textrm{if } e_{i-1}^{2^k}+e_1+1\not=0,
\end{array}\right.
\end{eqnarray}
where $e_i=\infty$ if $e_{i-1}^{2^k}+e_1+1=0$ and otherwise $e_i=\phi(e_{i-1}^{2^k})$ for any $i\ge 2$.

Suppose that \eqref{eq-2ik} holds for $i$. Then for the case of $i+1$ we have $x^{2^{(i+1)k}}=x^{2^k}=\phi(x)$ if $e_i=\infty$ and $e_{i+1}=\phi(e_i^{2^k})=\phi(\infty)=e_1$ which coincides with \eqref{eq-2ik}. If $e_i\not=\infty$, then by $x^{2^k}=\phi(x)=\frac{e_1x+1}{x+e_1+1}$ and \eqref{eq-2ik} we have
 \begin{eqnarray*}
 x^{2^{(i+1)k}}=\frac{e_i^{2^k}x^{2^k}+1}{x^{2^k}+e_i^{2^k}+1}=\frac{(e_1e_i^{2^k}+1)x+(e_i^{2^k}+e_1+1)}{(e_i^{2^k}+e_1+1)x+e_1e_i^{2^k}+e_i^{2^k}+e_1}.
\end{eqnarray*}
If $e_i^{2^k}+e_1+1=0$, then $e_1e_i^{2^k}+e_i^{2^k}+e_1=e_1e_i^{2^k}+1=e_1^2+e_1+1\ne 0$ as shown before. This means that $x^{2^{(i+1)k}}=x$ and $e_{i+1}=\infty$.
If $e_i^{2^k}+e_1+1\not=0$, then it can be readily verified that
\begin{eqnarray*}
 e_{i+1}=\frac{e_1e_i^{2^k}+1}{e_i^{2^k}+e_1+1}=\phi(e_i^{2^k}), \frac{e_1e_i^{2^k}+e_i^{2^k}+e_1}{e_i^{2^k}+e_1+1}=e_{i+1}+1.
\end{eqnarray*}
Therefore, \eqref{eq-2ik} holds for any $i\ge 2$.  This completes the proof. \done

\section{New Permutation Trinomials From Niho Exponents}

In this section, we present two new classes of permutation polynomials over $\ftwon$ of the form \eqref{def-PP}, namely
\begin{eqnarray*}
  f(x)=x+x^{s(2^m-1)+1}+x^{t(2^m-1)+1},
\end{eqnarray*}
where $n=2m$ and $1\leq s, t\leq 2^m$.

By using the property of the linear fractional polynomials over $\ftwon$, we can obtain new classes of permutation trinomials of the form \eqref{def-PP} as below.

\begin{Theorem}\label{thm-2k-1}
  Let $n=2m$ and $\gcd(2^k-1,2^m+1)=1$, where $m, k$ are positive integers with $k<m$. Then the trinomial $f(x)$ defined by \eqref{def-PP} is a permutation if $(s,t)=(\frac{2^k}{2^k-1},\frac{-1}{2^k-1})$.
\end{Theorem}

\Proof According to Lemma \ref{lem-Zieve}, to complete the proof, it is sufficient to prove that $h(x)=x(1+x^s+x^t)^{2^m-1}$ with $(s,t)=(\frac{2^k}{2^k-1},-\frac{1}{2^k-1})$ permutes the unit circle $U$ of $\ftwon$, which is equivalent to showing that $h(x^{2^k-1})=x^{2^k-1}(1+x^{s(2^k-1)}+x^{t(2^k-1)})^{2^m-1}=x^{2^k-1}(1+x^{2^k}+x^{-1})^{2^m-1}$ permutes $U$ since $\gcd(2^k-1,2^m+1)=1$. Notice that $x^{2^k+1}+x+1\not=0$ for any $x\in U$. Otherwise, from $x^{2^k+1}+x+1=0$ and $x^{2^m+1}+1=0$ we have $x^{2^k+1}+x+x^{2^m+1}=0$, i.e., $x^{2^m}=x^{2^k}+1$ which leads to $x=(x^{2^k}+1)^{2^m}=(x^{2^m}+1)^{2^k}=(x^{2^k})^{2^k}=x^{2^{2k}}$. Taking $2^k$-th power on both sides of $x^{2^k+1}+x+1=0$ gives $x^{2^{2k}+2^k}+x^{2^k}+1=0$. Then by $x^{2^{2k}}=x$ and $x^{2^k+1}+x+1=0$ we have $x^{2^k}=x$ which implies that $x=1$ due to $x^{2^k-1}=1$, $x^{2^m+1}=1$ and $\gcd(2^k-1,2^m+1)=1$, a contradiction. Thus, we arrive at $x^{2^k+1}+x+1\not=0$ for any $x\in U$, and then, $h(x^{2^k-1})=x^{2^k-1}(1+x^{2^k}+x^{-1})^{2^m-1}$ can be rewritten as $h(x^{2^k-1})=\frac{x^{2^k+1}+x^{2^k}+1}{x^{2^k+1}+x+1}$.

We next prove that $h(x^{2^k-1})=\frac{x^{2^k+1}+x^{2^k}+1}{x^{2^k+1}+x+1}$ permutes the unit circle of $\ftwon$. For any given $a\in U$, we show that $h(x^{2^k-1})=a$ has at most one solution in $U$. Note that $h(x^{2^k-1})=a$ can be expressed as
 \begin{eqnarray}\label{eq-LFP}
x^{2^k}=\frac{ax+(a+1)}{(a+1)x+1}.
\end{eqnarray}
We then can discuss it as follows:
\begin{enumerate}
  \item [1)] $a=1$. If $a=1$, then \eqref{eq-LFP} is reduced to $x^{2^k}=x$ which has exactly one solution $x=1$ in $U$ since $\gcd(2^k-1,2^m+1)=1$.
  \item [2)] $a=(a+1)^2$, i.e., $a^2+a+1=0$. This case happens only if $m$ is odd since $a\in U$.  Then, \eqref{eq-LFP} can be reduced to $x^{2^k}=a+1=a^2$ which has exactly one solution in $U$ when this case happens.
  \item [3)] $a^3\not=1$. For this case, if \eqref{eq-LFP} holds for some $x\in U$, then by Lemma \ref{lem-phi}, we have that \eqref{eq-2ik} holds for the same $x$ for any $i\ge 2$. If $k$ is odd, then $x^{2^{ik}}=\frac{1}{x}$ if one takes $i=m$. Thus, by \eqref{eq-2ik} we have either $x=\frac{1}{x}$ or $\frac{e_mx+1}{x+e_m+1}=\frac{1}{x}$, it then can be readily seen that no matter which case there exists at most one such $x$. If $k$ is even, assume that $k=2^jk_1$ for some positive $j$ and odd $k_1$, then we can conclude that $2^j|m$ since $\gcd(2^k-1,2^m+1)=1$ if and only if $\frac{k}{\gcd(m,k)}$ is odd. Take $i=\frac{m}{2^j}$ in \eqref{eq-2ik}, we have  $x^{2^{ik}}=x^{2^{mk_1}}=\frac{1}{x}$ since $k_1$ is odd. Therefore, the proof can be completed as the case of $k$ is odd.
\end{enumerate}
This completes the proof. \done

If one takes $k=1$, then Theorem \ref{thm-2k-1} generalizes Theorem 3.2 in \cite{Ding-QWYY}.

\begin{Corollary}\label{cor-k=1}
   Let $n=2m$ for a positive integer $m$. Then the trinomial $f(x)$ defined by \eqref{def-PP} is a permutation if $(s,t)=(2,-1)=(2,2^m)$.
\end{Corollary}

If one takes $k=2$, then Theorem \ref{thm-2k-1} generalizes Theorem 3 in \cite{LH}.

\begin{Corollary}\label{cor-k=2}
   Let $n=2m$ for an even integer $m$. Then the trinomial $f(x)$ defined by \eqref{def-PP} is a permutation if $(s,t)=(\frac{4}{3},\frac{-1}{3})=(\frac{2^m+5}{3},\frac{2^{m+1}+1}{3})$.
\end{Corollary}

\begin{Theorem}\label{thm-2k+1}
  Let $n=2m$ and $\gcd(2^k+1,2^m+1)=1$, where $m, k$ are positive integers. Then the trinomial $f(x)$ defined by \eqref{def-PP} is a permutation if $(s,t)=(\frac{1}{2^k+1},\frac{ 2^k}{2^k+1})$.
\end{Theorem}

\Proof According to Lemma \ref{lem-Zieve}, we need to show that $h(x)=x(1+x^s+x^t)^{2^m-1}$ permutes the unit circle $U$ of $\ftwon$ if $(s,t)=(\frac{1}{2^k+1},\frac{ 2^k}{2^k+1})$. Since $\gcd(2^k+1,2^m+1)=1$, then it suffices to prove that $h(x^{2^k+1})=x^{2^k+1}(1+x^{s(2^k+1)}+x^{t(2^k+1)})^{2^m-1}=x^{2^k+1}(1+x+x^{2^k})^{2^m-1}$ permutes $U$. First, we show that $x^{2^k}+x+1\not=0$ for any $x\in U$. Suppose that $x^{2^k}+x+1=0$ and $x^{2^m+1}+1=0$, then one gets $x^{2^k}=x+1$ and $x^{2^m}=\frac{1}{x}$ which leads to $x^{2^{m+k}}=(x+1)^{2^m}=(\frac{1}{x})^{2^k}$. This together with $x^{2^k}+x+1=0$ implies that $(x+1)^{2^m}=\frac{1}{x+1}$, i.e., $(x+1)^{2^m+1}=1$, and then, we obtain that $x^2+x+1=0$ due to $x^{2^m+1}+1=0$. However, this is impossible if $m$ is even since $x^2+x+1=0$ has no solution in $U$ for even $m$. If $m$ is odd, then $x^2+x+1=0$ has two solutions in $U$ satisfying $x^3=1$. Note that $\gcd(2^k+1,2^m+1)=1$ if and only if one of $\frac{m}{\gcd(m,k)}$ and $\frac{k}{\gcd(m,k)}$ is even which means that $k$ is even and $2^k\equiv 1\pmod{3}$. This leads to $x^{2^k}+x+1=x+x+1=1$, a contradiction. Hence, $x^{2^k}+x+1\not=0$ for any $x\in U$, and $h(x^{2^k+1})=x^{2^k+1}(1+x+x^{2^k})^{2^m-1}$ can be expressed as $h(x^{2^k+1})=\frac{x^{2^k+1}+x^{2^k}+x}{x^{2^k}+x+1}$.

We next prove that $h(x^{2^k+1})=\frac{x^{2^k+1}+x^{2^k}+x}{x^{2^k}+x+1}$ permutes the unit circle of $\ftwon$. For any given $a\in U$, we show that $h(x^{2^k+1})=a$ has at most one solution in $U$. Note that $h(x^{2^k+1})=a$ can be written as
\begin{eqnarray*}
x^{2^k}=\frac{(a+1)x+a}{x+(a+1)}.
\end{eqnarray*}
Taking $2^m$-th power on both sides of the above equation gives
 \begin{eqnarray}\label{eq-LFP-2k+1}
x^{2^{m+k}}=\frac{(\overline{a}+1)\overline{x}+\overline{a}}{\overline{x}+(\overline{a}+1)}=\frac{\overline{a}x+(\overline{a}+1)}{(\overline{a}+1)x+1}.
\end{eqnarray}

Let $k'=m+k$ and $a'=\overline{a}$. Thus, by Lemma \ref{lem-phi}, if $a'^3\not=1$, then for any $i\ge 2$ we have
 \begin{eqnarray}\label{eq-2ik+1}
 x^{2^{ik'}}=\left \{ \begin{array}{cl}
x, & \textrm{if } {e'}_{i-1}^{2^{k'}}+e'_1+1=0, \\
\frac{e'_ix+1}{x+e'_i+1}, & \textrm{if } {e'}_{i-1}^{2^{k'}}+e'_1+1\not=0,
\end{array}\right.
\end{eqnarray}
where $e'_i=\infty$ if ${e'}_{i-1}^{2^{k'}}+e'_1+1=0$ and otherwise $e'_i=\phi({e'}_{i-1}^{2^{k'}})$ for any $i\ge 2$. Thus, similar as the proof of Theorem \ref{thm-2k-1}, we can discuss it as follows:
\begin{enumerate}
  \item [1)] $a'=1$. If $a'=1$, then \eqref{eq-LFP-2k+1} is reduced to $x^{2^{m+k}}=x$, i.e., $x^{2^k+1}=1$,  which has exactly one solution $x=1$ in $U$ since $\gcd(2^k+1,2^m+1)=1$.
  \item [2)] $a'=(a'+1)^2$, i.e., $a'^2+a'+1=0$. This case happens only if $m$ is odd since $a'\in U$.  Then, \eqref{eq-LFP-2k+1} can be reduced to $x^{2^{m+k}}=a'+1=a'^2$ which has exactly one solution in $U$ when this case happens.
  \item [3)] $a'^3\not=1$. For this case, if \eqref{eq-LFP-2k+1} holds for some $x\in U$, then by Lemma \ref{lem-phi}, we have that \eqref{eq-2ik+1} holds for the same $x$ for any $i\ge 2$. If $k'$ is odd, then $x^{2^{ik'}}=\frac{1}{x}$ if one takes $i=m$. Thus, by \eqref{eq-2ik+1} we have either $x=\frac{1}{x}$ or $\frac{e'_mx+1}{x+e'_m+1}=\frac{1}{x}$, which implies that there exists at most one such $x$. If $k'$ is even, then we can claim that $m$ is even. Moreover, let $k'=2^{j'}k'_1$ for some positive $j'$ and odd $k'_1$, we have $2^{j'}|m$. This can be verified by the fact that $\gcd(2^k+1,2^m+1)=1$ if and only if one of $\frac{m}{\gcd(m,k)}$ and $\frac{k}{\gcd(m,k)}$ is even, i.e., $\frac{m}{\gcd(m,k)}+\frac{k}{\gcd(m,k)}=\frac{m+k}{\gcd(m,k)}=\frac{k'}{\gcd(m,k)}$ is odd. Therefore, for even $k'=2^{j'}k'_1$, we can take $i=\frac{m}{2^{j'}}$ in \eqref{eq-2ik+1} and then $x^{2^{ik'}}=x^{2^{mk'_1}}=\frac{1}{x}$ since $k'_1$ is odd. Then, the proof can be completed as the case of $k'$ is odd.
\end{enumerate}
This completes the proof. \done

If one takes $k=1$, then Theorem \ref{thm-2k+1} shows that the trinomial $f(x)$ defined by \eqref{def-PP} is a permutation for even $m$ if $(s,t)=(\frac{1}{2^k+1},\frac{ 2^k}{2^k+1})=(\frac{1}{3},\frac{2}{3})$. However, this result is covered by Theorem 3.4 in \cite{Ding-QWYY} up to equivalence (see Table \ref{Table1}).

If one takes $k=2$, then Theorem \ref{thm-2k+1} generalizes Theorem 6 in \cite{LH}.

\begin{Corollary}\label{cor-k+=2}
  Let $n=2m$ satisfy $\gcd(5,2^m+1)=1$. Then the trinomial $f(x)$ defined by \eqref{def-PP} is a permutation if $(s,t)=(\frac{1}{5},\frac{4}{5})$.
\end{Corollary}

\begin{table}[ht]
\caption{Known pairs $(s,t)$ such that $f(x)$ defined by \eqref{def-PP} are permutation polynomials}\label{Table1}
\begin{center}{
\begin{tabular}{|c|c|c|c|c|c|}
  \hline
No. &  $(s, t)$ & $h(x)$& Conditions & Equivalent Pairs& Proven in  \\ \hline\hline
 1  & $(k, -k)$ & $x$ & see Thm. 3.4 in \cite{Ding-QWYY}   &  $(\frac{\pm k}{2k\mp1},\frac{\pm2k}{2k\mp1})$ & \cite{Ding-QWYY} \\ \hline
2 &  $(2, -1)$ & $\frac{x^3+x^2+1}{x^3+x+1}$&  positive $m$  & $(1,\frac{1}{3})$, $(1,\frac{2}{3})$ & \cite{Ding-QWYY}  \\ \hline
3  & $(1, -\frac{1}{2})$& $\frac{x(x^3+x^2+1)}{x^3+x+1}$ &  $\gcd(3,m)=1$  & $(1,\frac{3}{2})$, $(\frac{1}{4},\frac{3}{4})$ & \cite{Li-Qu-Chen,LQLF,GS} \\ \hline
4  &  $(-\frac{1}{3}, \frac{4}{3})$& $\frac{x^5+x^4+1}{x^5+x+1}$ &  $m$ even & $(1,\frac{1}{5})$, $(1,\frac{4}{5})$ &  \cite{LH,LQLF}  \\ \hline
5  &   $(3, -1)$& $\frac{x^4+x^3+1}{x(x^4+x+1)}$ &  $m$ even & $(\frac{3}{5},\frac{4}{5})$, $(\frac{1}{3},\frac{4}{3})$ &  \cite{LH,LQLF}   \\ \hline
 6  &   $(-\frac{2}{3}, \frac{5}{3})$& $\frac{x^7+x^5+1}{x^7+x^2+1}$ &  $m$ even & $(1,\frac{2}{7})$, $(1,\frac{5}{7})$ &  \cite{LH}   \\ \hline
7  &     $(\frac{1}{5}, \frac{4}{5})$ & $\frac{x(x^4+x^3+1)}{x^4+x+1}$&  $\gcd(5,2^m+1)=1$ & $(1,-\frac{1}{3})$, $(1,\frac{4}{3})$ &  \cite{LH,LQLF,GS}  \\ \hline
8  &  $(2, -\frac{1}{2})$ & $\frac{x(x^5+x+1)}{x^5+x^4+1}$ &  $m\equiv 2,4\pmod{6}$&  $(\frac{2}{3},\frac{5}{6}), (\frac{1}{4},\frac{5}{4})$ &  \cite{LQLF}\\ \hline
9  &  $(4, -2)$ & $\frac{x^6+x^4+1}{x(x^6+x^2+1)}$ & $\gcd(3,m)=1$  &  $(\frac{2}{3},\frac{5}{6}), (\frac{1}{4},\frac{5}{4})$ & \cite{LQLF} \\ \hline
10  &   $(\frac{2^k}{2^k-1},\frac{-1}{2^k-1})$& $\frac{x^{2^k+1}+x^{2^k}+1}{x^{2^k+1}+x+1}$  & $\gcd(2^k-1,2^m+1)=1$   &  $(1,\frac{1}{2^k+1})$, $(1,\frac{2^k}{2^k+1})$ & Theorem \ref{thm-2k-1} \\ \hline
11 &   $(\frac{1}{2^k+1},\frac{2^k}{2^k+1})$ & $\frac{x^{2^k+1}+x^{2^k}+x}{x^{2^k}+x+1}$ & $\gcd(2^k+1,2^m+1)=1$  &  $(1,\frac{2^k}{2^k-1})$, $(1,\frac{-1}{2^k-1})$ & Theorem \ref{thm-2k+1} \\  \hline
\end{tabular}}
\end{center}
\end{table}

To end this section, we list all the known pairs $(s,t)$ such that the polynomials of the form \eqref{def-PP} are permutations in Table \ref{Table1} and we claim that all the permutation trinomials obtained in this paper are multiplicative inequivalent with the known ones. Notice that the inverse of a normalized Niho exponent, if exists, is still a normalized Niho exponent and the product of two normalized Niho exponents is also a normalized Niho exponent. With this fact, it can be readily checked that the permutation trinomials obtained in this paper are multiplicative inequivalent with the known permutation trinomials listed in Theorem 1 and Table 1 in \cite{LH}. Theorem 1 in \cite{LH} listed all the known permutation trinomials over $\ftwon$ for even $n$ and Table 1 in \cite{LH} presented all the known permutation trinomials of the form \eqref{def-PP}.  The ``Equivalent Pairs" column in Table \ref{Table1} are obtained based on Lemma 4 in \cite{LH}, which also leads to permutation trinomials of the form \eqref{def-PP} if they exist.  Note that  the fractional polynomial in No. 10 covers those of in Nos. 2 and 4 as special cases and the fractional polynomial in No. 11 covers that of in No. 7 as a special case.

Further,  it also should be noted that trinomial permutations over $\ftwon$ with a more general form $f_{r,s,t}(x)=x^r+x^{s(2^m-1)+r}+x^{t(2^m-1)+r}$ can also be obtained by using the same techniques as in \cite{LH} and this paper if the parameters $r,s,t$ are suitably chosen based on Lemma \ref{lem-Zieve}.  Lemma \ref{lem-Zieve} implies that the determination of the permutation property of $f_{r,s,t}(x)$ over $\ftwon$ is equivalent to that of the permutation property of $h_{r,s,t}(x)=x^r(1+x^s+x^t)^{2^m-1}=\frac{x^r+x^{r-s}+x^{r-t}}{1+x^s+x^t}$ over the unit circle of $\ftwon$.  From the viewpoint of Lemma \ref{lem-Zieve}, the trinomials   $f_{r_1,s_1,t_1}(x)$ and $f_{r_2,s_2,t_2}(x)$ are viewed as the same in Table \ref{Table1} for different integer tuples  $(r_1,s_1,t_1)$ and $(r_2,s_2,t_2)$ if  $h_{r_1,s_1,t_1}(x)= h_{r_2,s_2,t_2}(x)$  although $f_{r_1,s_1,t_1}(x)$ and $f_{r_2,s_2,t_2}(x)$ are multiplicative inequivalent (anyway one of them can be easily obtained from the other).  With this observation, the two conjectures proposed in \cite{GS} can be settled by using the fifth and fourth fractional permutation polynomials in Table \ref{Table1} over the unit circle of $\ftwon$ proved in \cite{LH,LQLF} .

\section{Conclusion Remarks}

In this paper new classes of permutation trinomials over $\ftwon$ with the form \eqref{def-PP} were obtained from Niho exponents by using the property of linear fractional polynomials and some techniques in solving equations over finite fields. It was shown that the presented results generalized some earlier works in \cite{Ding-QWYY,GS,LH,LQLF}. It is interesting to generalize our method in general or find new ideas to prove the permutation property of fractional polynomials over finite fields in order to obtain more permutation trinomials.

%\section*{Acknowledgements}

\end{document}